\documentstyle[12pt]{article}
\newlength{\mathspace}
\def\np#1{ Nucl. Phys. B#1}
\def\pr#1    { Phys. Rev. D#1 }
\def\pl#1{ Phys. Lett. B#1}
\def\cmp   { Commun. Math. Phys. }

\def\ijmp#1  { Int. Jour. Mod. Phys. A#1 }
\def\mpl#1   { Mod. Phys. Lett. A#1 }
\def\del {\partial}
\def\begineq{\begin{equation}}
\def\endeq{\end{equation}}
\def\eqabegin{\begin{eqnarray}}
\def\eqaend{\end{eqnarray}}
\def\nn{\nonumber}

\def\parbigskip        {  \par\bigskip  }
\def\parmedskip        {  \par\medskip  }

\def\parbigskipn        {  \par\bigskip\noindent  }

\begin{document}
\baselineskip=0.7cm
\setlength{\mathspace}{2.5mm}

%%%%%%%%%%%%%%%%%%%%%%%%%%%%%%%%%%%%%%%%%%%%%%%%%%%%%%%%%%

                                %titlepage

%%%%%%%%%%%%%%%%%%%%%%%%%%%%%%%%%%%%%%%%%%%%%%%%%%%%%%%%%%
\begin{titlepage}

    \begin{normalsize}
     \begin{flushright}
                 UG-9/94 \\
                 hep-th/9410233 \\
     \end{flushright}
    \end{normalsize}
    \begin{LARGE}
       \vspace{1cm}
       \begin{center}
         $c_M<1$ String Theory as a Constrained \\
         Topological Sigma Model\\
       \end{center}
    \end{LARGE}

  \vspace{5mm}

\begin{center}
           Pablo M. L{\sc latas}
           \footnote{E-mail address:
              llatas@th.rug.nl}
           \footnote{Address after April 1, 1995: Department of Physics,
                    University of California at Santa Barbara, CA 93106, USA.}
                \ \  and \ \
           Shibaji R{\sc oy}
           \footnote{E-mail address:
              roy@th.rug.nl}
           \footnote{Address after January 15, 1995: Departamento de Fisica
                     de Particulas, Universidade de Santiago, E-15706,
                     Santiago de Compostela, Spain.}\\
      \vspace{4mm}
        {\it Institute for Theoretical Physics} \\
        {\it Nijenborgh 4, 9747 AG Groningen}\\
        {\it The Netherlands}\\
      \vspace{1cm}

    \begin{large} ABSTRACT \end{large}
        \par
\end{center}
\begin{quote}
 \begin{normalsize}
\ \ \ \
It has been argued by Ishikawa and Kato that by making use of a specific
bosonization, $c_M=1$ string theory can be regarded as a constrained
topological sigma model. We generalize their construction for any $(p,q)$
minimal model coupled to two dimensional (2d) gravity and show that the
energy--momentum tensor and the topological charge of a constrained
topological sigma model can be mapped to the energy--momentum tensor and
the BRST charge of $c_M<1$ string theory at zero cosmological constant.
We systematically study the physical state spectrum of this
topological sigma model and recover the spectrum in the absolute cohomology
of $c_M<1$ string theory. This procedure provides us a manifestly topological
representation of the continuum Liouville formulation of $c_M<1$ string
theory.
 \end{normalsize}
\end{quote}

\end{titlepage}
\vfil\eject
It has been shown recently from various points of view that $c_M=1$ string
theory has manifestly topological field theoretic descriptions. It was first
pointed out in ref.[1] that a special Kazama-Suzuki coset model is equivalent
to $c_M=1$ matter coupled to 2d gravity. Further arguments in favor of the
topological nature of $c_M=1$ string theory were given by identifying it with
a topological sigma model [2], a topological $G/G$ model [3] as well as a
topological
Landau-Ginzburg model [4,5] with a particular superpotential. The latter
identification also clarified the origin of the long suspected integrability
structure [5,6] in $c_M=1$ string theory. It should be pointed out here
that in
these works the equivalence was established by comparing the cohomology
structure as well as by computing some correlation functions which agree with
the matrix--model results. A more direct approach, clarifying the reason why
the observables can be obtained from a topological model, was taken by
Ishikawa and Kato in ref.[7]. They have shown that by making use of a specific
bosonization one can identify $c_M=1$ string theory with a topological sigma
model at the level of Lagrangians rather than at the level of amplitudes.

The topological nature of the Liouville approach to $c_M<1$ string theory
is not as clear as in $c_M=1$ case. It has long been known that certain
topological matter coupled to 2d topological gravity reproduce [8,9] the
matrix--model results of $c_M<1$ string theory. The Landau-Ginzburg formulation
and the integrability structure in this case are also fairly
well-understood [10].
After a considerable amount of effort a family of twisted $N=2$ superconformal
structures have been revealed [11,12,13] in the continuum Liouville approach
to $c_M<1$
string theory indicating a close relationship with some topological field
theories. Using this information it became clear why $(1,q)$ models coupled
to gravity are topological [12]. But still a manifestly topological
representation
of the Liouville formulation of $c_M<1$ string theory remained illusive. An
attempt in this direction was made in ref.[14]. By using a bosonization (which
reduced to the topological gravity formulation of Distler [15] as a special
case),
we found that there is a topological gravity structure in any $(p,q)$ model
coupled
to 2d gravity. It was noted also that the total BRST charge of the topological
gravity is different from the string BRST charge and hence it was not clear
how to obtain the full spectrum of $c_M<1$ string theory in the topological
gravity representation.

In this paper, we look at a different bosonization similar to the one found
in ref.[7] for $c_M=1$ string theory. We generalize the construction for any
$(p,q)$ minimal model coupled to 2d gravity and show in analogy that $c_M<1$
string theory can also be regarded as a topological sigma model [17] where
one of
the coordinates is identified with a ground ring generator [18,19].
In particular, we
show that the energy--momentum tensor and the topological charge of the
topological sigma model can be mapped to the energy--momentum tensor and the
BRST charge of $c_M<1$ string theory at zero cosmological constant.
This approach also clarifies the origin of the twisted $N=2$
superconformal algebra in $c_M<1$ string theory. We have also systematically
studied the physical state spectrum of this topological sigma model and found
that they coincide with the spectrum in the absolute cohomology of $c_M<1$
string theory. A detailed description will be presented elsewhere [20].

The $(p,q)$ minimal models (where gcd ($p,q$)=1) coupled to 2d gravity can be
described in terms of Coulomb gas representation with the total
energy--momentum tensor,
\begineq
T(z) \,=\, T_M(z) + T_L(z) + T^{gh}(z)
\endeq
where the matter, Liouville and the ghost energy--momentum tensors are given
respectively as,
\eqabegin
T_M(z) &=& -{1\over 2}:\del\phi_M(z)\del\phi_M(z): + i Q_M \del^2\phi_M(z) \\
T_L(z) &=& -{1\over 2}:\del\phi_L(z)\del\phi_L(z): + i Q_L \del^2\phi_L(z) \\
T^{gh}(z) &=& -2 :b(z)\del c(z): - :\del b(z) c(z):
\eqaend
Here we are working in a free theory with zero cosmological constant and
so, we
concentrate only in the holomorphic sector. Here $\phi_M(z)$, $\phi_L(z)$
denote the matter and Liouville field with the propagators having the form
$\langle \phi_M(z) \phi_M(w)\rangle = \langle \phi_L(z) \phi_L(w)\rangle =
-\log (z-w)$.
$2Q_M$, $2Q_L$ denote the background charges for the matter and Liouville
sector which satisfy $Q_M^2 + Q_L^2 = -2$, since the total central charge
of the combined matter-Liouville theory is 26. $(b(z), c(z))$ are the usual
reparametrization ghost system having conformal weights 2 and $-1$ respectively
with the operator product expansion (OPE) given as $b(z) c(w) \sim \frac
{1}{z-w}$. Since the $(p,q)$ minimal models are characterized by the Virasoro
central charge $1-\frac{6(p-q)^2}{pq}$, the background charges for the matter
and Liouville sector can be parametrized as,
\eqabegin
Q_M &=& \left(\frac {1}{2\lambda}-\lambda\right)\\
Q_L &=& i \left(\frac {1}{2\lambda}+\lambda\right)
\eqaend
where $\lambda = \sqrt{\frac {q}{2p}}$. We now define
the following four conformal fields,
\eqabegin
x(z) &=& :\left[b(z)c(z)-\frac{i}{2\lambda}\Big(\del\phi_M(z) + i\del\phi_L(z)
\Big)
\right]\,e^{i\lambda\big(\phi_M(z)-i\phi_L(z)\big)}:\\
\bar{p}(z) &=& :e^{-i\lambda\big(\phi_M(z) - i\phi_L(z)\big)}:\\
B(z) &=& :b(z)\,e^{i\lambda\big(\phi_M(z) - i\phi_L(z)\big)}:\\
C(z) &=& :c(z)\,e^{-i\lambda\big(\phi_M(z) - i\phi_L(z)\big)}:
\eqaend
with conformal weights 0, 1, 1, 0 respectively with respect to (1). One can
easily verify that the OPEs among these fields are,
\begineq
\begin{array}{rcccl}
\bar{p}(z) x(w) &=& - x(z) \bar{p}(w) &\sim& {\displaystyle\frac{1}{z-w}}
\\ [\mathspace]
B(z) C(w) &=& C(z) B(w) &\sim& {\displaystyle\frac{1}{z-w}}
\end{array}
\endeq
with the rest of the OPEs being regular. We note that the energy--momentum
tensor (1) is symmetric under the interchange $\lambda\leftrightarrow \frac{1}
{2\lambda}$, $\phi_M(z)\leftrightarrow-\phi_M(z)$ and $\lambda \leftrightarrow
\frac{1}{2 i \lambda}$, $\phi_M(z)\leftrightarrow \phi_L(z)$ and so, there
exists another bosonization which can be obtained by using this symmetry.
In terms of these fields (7-10) the
energy--momentum tensor (1) turns out to be,
\begineq
T(z) = -:\del x(z) \bar{p}(z): - :B(z) \del C(z):
\endeq
By identifying $\bar{p} \equiv \del \bar{x}$, where $\bar{x}$ denotes
the holomorphic part of the
complex conjugate of the coordinate $x$, we easily recognize that the
energy--momentum
tensor (12) can be obtained from the topological sigma model (in one
complex dimension) action [17].
We also
point out that for $\lambda = \frac{1}{\sqrt 2}$, the value for $c_M$=1
string theory, and $\phi_L(z) \rightarrow  -\phi_L(z)$ our bosonization (7-10)
matches precisely with Eq.(2.3) in ref.[7]. We would like to mention here
that one of the coordinates of the topological sigma model Eq.(7) is nothing
but one
of the ground ring generators of $c_M<1$ string theory. This is a true
conformal field of weight zero in the sense that it does not contain the
$\log z$ term in its expansion in $z$ like a proper sigma model
coordinate. In this sense, the topological sigma model in question here is
a constrained (zero momentum sector) one.

The nilpotent supersymmetry current of this topological sigma model has the
form
\begineq
Q(z) = - C(z) \del x(z)
\endeq
Substituting (10) and (7) we find
\eqabegin
Q(z) &=& :c(z)\left[T_M(z) + T_L(z) + \frac {1}{2} T^{gh}(z)\right]: \nn \\
     & &\qquad +\frac{1}{2}\del\left[\del c(z) + 2\lambda c(z) \bigg(
i \del\phi_M(z) + \del\phi_L(z)\bigg)\right]
\eqaend
Therefore, we conclude from (14), that the topological charge of the
topological sigma
model is the same as the BRST charge of $c_M<1$ string theory. The origin of
the twisted $N=2$ superconformal algebra in $c_M<1$ string theory [12,13]
can now be
understood in terms of the corresponding algebra in the associated topological
sigma model. It is straightforward to check that the following generators
of the topological sigma model
\eqabegin
T(z) &=& :\del x(z) \bar{p}(z): - :B(z)\del C(z):\\
G^+(z) &=& C(z) \del x(z) - (a_3 - \frac {1}{2})\del \big(C(z) x(z)\big)\\
G^-(z) &=& B(z) \bar{p}(z)\\
J(z) &=& :C(z) B(z): + (a_3 -\frac{1}{2})\Big[:C(z) B(z): + :x(z) \bar{p}(z)
:\Big]
\eqaend
satisfy a twisted superconformal algebra with the associated $N=2$ central
charge $c^{N=2} = 6a_3$, where $a_3$ is a constant parameter.
We note that $G^+(z)$ is
the topological current (13) modified by a total derivative term, so that, the
topological charge is not affected. Also, the $U(1)$ current $J(z)$ has been
modified suitably so that they form a closed algebra. For $a_3 = \frac{1}{2}$
these generators were described in the context of topological sigma models
in [21].
Substitution of
the topological sigma model fields in terms of the fields in string theory
(7-10),
gives the corresponding generators in $c_M<1$ string theory,
\eqabegin
T(z) &=& -{1\over 2}:\del\phi_M(z)\del\phi_M(z): + i Q_M \del^2\phi_M(z)
         -{1\over 2}:\del\phi_L(z)\del\phi_L(z): + i Q_L \del^2\phi_L(z) \nn\\
     & &\qquad\qquad\qquad-2:b(z)\del c(z): - :\del b(z) c(z):\\
G^+(z) &=& :c(z)\left[T_M(z) + T_L(z) + \frac{1}{2} T^{gh}(z)\right]: \nn \\
       & &\qquad\qquad +a_1 \del\Big(c(z)\del\phi_L(z)\Big) + a_2
\del\Big(c(z)\del \phi_M(z)\Big)
+a_3 \del^2 c(z)\\
G^-(z) &=& b(z)\\
J(z) &=& :c(z) b(z): -a_1 \del\phi_L(z) - a_2 \del\phi_M(z)
\eqaend
where $a_1 = \frac{1}{4}\big[i Q_L(2a_3-3) - Q_M(2a_3+1)\big]$ and
$a_2 = \frac{1}{4}
\big[i Q_M(2a_3-3) + Q_L(2a_3+1)\big]$. This formalism, therefore, clarifies
the origin
of the twisted $N=2$ superconformal algebra in $c_M<1$ string theory found in
ref.[13].

In order to find the physical state spectrum, we note that the topological
charge
of the topological sigma model is given by\footnote[1]{A conformal field of
weight $h$ is expanded in terms of modes as $\phi(z) = \sum_n \phi_n\,z^{-n-h}
$.}
\eqabegin
Q &=& -\oint dz\, C(z)\del x(z) \nn \\
  &=& \sum_m m C_{-m} x_m
\eqaend
Physical states of the topological sigma model are the states which are in
the kernel of this charge modulo its image. First we note that the
vacuum of the topological sigma model is characterized by the following
regularity conditions,
\begineq
\begin{array}{rcccl}
B_n |0\rangle_t &=& \bar{p}_n |0\rangle_t &=& 0 \qquad\qquad
{\rm for}\qquad n \geq 0 \\[\mathspace]
C_n |0\rangle_t &=& x_n |0\rangle_t &=& 0 \qquad\qquad {\rm for}
\qquad n \geq 1
\end{array}
\endeq
which follow from their respective conformal weights. Using (24) it is easy to
check that $|0\rangle_t$ is physical with respect to (23). Since
the topological
charge (23) is the same as the string BRST charge we identify $|0\rangle_t$ as
the $SL(2,{\bf C})$ invariant vacuum and so we drop the subscript `$t$'
subsequently. The physical spectrum of the topological sigma model can be built
by applying the physical modes on this vacuum. In order to determine the
physical modes we note that,
\eqabegin
\left[Q , x_n\right] &=& 0 \\
\left[Q , \bar{p}_n\right] &=& -n\, C_n \\
\{Q , B_n\} &=& n\, x_n \\
\{Q , C_n\} &=& 0
\eqaend
where $[\,\,\,]$ and $\{\,\,\,\}$ indicate commutator and anticommutator
respectively.
It is clear from (25-28) that for $n \neq 0$, all the modes are non-physical as
they form ``quartets'', but for $n=0$ this is no longer true and $x_0$,
$\bar{p}_0$, $B_0$ and $C_0$ are the physical modes. Since $B_0$ and $p_0$
annihilate the topological vacuum, the physical states would have the form
$C_0^\epsilon x_0^n |0\rangle$, where $\epsilon = 0$ or 1 and $n=0,1,2,
\ldots $. After a few simple calculations we identify these states with
some of the
states of $c_M<1$ string theory [19,22,23,24,25] as follows,
\eqabegin
x_0^n |0\rangle &= & :x^n (0): |0\rangle \\
C_0 |0\rangle &= & :c(0) e^{-i\lambda\big(\phi_M(0)-i\phi_L(0)\big)}:
|0\rangle \\
C_0 x_0 |0\rangle &= & -\left(\del c(0) - \frac{1}{2\lambda}c(0)
\del\phi_L(0) + \frac{i}{2\lambda}c(0)\del\phi_M(0)\right) |0\rangle \nn \\
              &= & a(0) |0\rangle \\
\eqaend
and in general
\begineq
C_0 x_0^n |0\rangle = :a(0) x^{n-1}(0): |0\rangle
\endeq
We notice that the physical state (30) is a tachyonic state whose matter
momentum lies on the edge of the Kac-table. Also, the state $a(z) =
\left[Q, \frac {1}{2\lambda}\Big(\phi_L(z) - i \phi_M(z)\Big)\right] =$
\newline $-\left(\del c(z) - \frac{1}{2\lambda}c(z)\del\phi_L(z) +
\frac{i}{2\lambda}
c(z)\del\phi_M(z)\right)$ is a physical state in the absolute cohomology [18]
since $\oint dz\,z b(z) a(0) \neq 0$. So, in this way we recover part of
the spectrum
of $c_M<1$ string theory in the absolute cohomology.

In order to obtain the rest of the physical states we recall [26] that
in a bosonization, there are more distinct inequivalent representations of
vacua in the original theory, known as the picture changed vacua, which should
be included in order to compare the physical states of both theories.
Since $c_M<1$ string theory is a
bosonized form of the topological sigma model, we recover the rest of the
states from the picture changed vacua.
The most general picture changed vacuum can be obtained
from the bosonization formula (7-10) and has the form (details will be
given in [20]),
\eqabegin
|q_1, q_2\rangle &=& : e^{i(q_1 Q_M -\lambda q_2)\phi_M(0) + i (q_1 Q_L
+i\lambda q_2)\phi_L(0) - i(q_1 + q_2)\psi(0)}: |0\rangle \nn \\
                 &=& :e^{i\big(\frac{q_1}{2\lambda}-(q_1+q_2)\lambda\big)
\phi_M(0)
-\big(\frac{q_1}{2\lambda}+(q_1+q_2)\lambda\big)\phi_L(0)-i(q_1+q_2)\psi(0)}:
|0\rangle
\eqaend
Here $q_1$ and $q_2$ are some fixed numbers. $\psi(z)$ is a bosonic field
obtained
from the bosonization of the reparametrization ghosts $b(z) =
:e^{i \psi(z)}:$ and $c(z) = e^{- i \psi(z)}:$. In terms of modes the picture
changed vacuum satisfies
\eqabegin
x_n |q_1, q_2\rangle &=& 0 \qquad\qquad {\rm for}\qquad n\geq 1-q_1\\
\bar{p}_n |q_1, q_2\rangle &=& 0 \qquad\qquad {\rm for}\qquad n\geq q_1\\
B_n |q_1, q_2\rangle &=& 0 \qquad\qquad {\rm for}\qquad n\geq q_2\\
C_n |q_1, q_2\rangle &=& 0 \qquad\qquad {\rm for}\qquad n\geq 1-q_2
\eqaend
Using (35-38) we find that the topological charge (23) acting
on the picture changed vacuum gives,
\eqabegin
Q |q_1, q_2\rangle &=& \sum_m m\,C_{-m} x_m |q_1, q_2\rangle \nn\\
        &=& -\sum_{m=q_1}^{m=-q_2} m\, x_{-m} C_m |q_1, q_2\rangle
\eqaend
So, the vacuum is $Q$-invariant if $q_1+q_2 \geq 1$. When $q_1+q_2
<1$ an invariant vacuum could be constructed as $C_{-q_2} C_{-q_2-1}
\ldots C_{q_1+2} C_{q_1+1} C_{q_1} |q_1, q_2\rangle$ since $C$ is a
fermionic field. By using $C_n = \oint dz\, z^{n-1} C(z)$ and Eq.(10),
we can easily show that,
\begineq
C_{-q_2} C_{-q_2-1} \ldots C_{q_1+2} C_{q_1+1} C_{q_1} |q_1, q_2\rangle
= |q_1, 1-q_1\rangle
\endeq
So, the most general $Q$-invariant picture changed vacuum has the form
\begineq
|q_1, q_2\rangle \qquad\qquad {\rm where} \qquad q_1+q_2 \geq 1
\endeq
and the vacuum (40) is already contained in (41) as a special case.
We also note
that in order for this vacuum to be non-trivial (non $Q$-exact), its conformal
weight would have to be zero. Since the conformal weight of the picture
changed vacuum $|q_1, q_2\rangle$ is $\frac {1}{2} (q_2-q_1)(q_1+q_2-1)$ this
will happen only when $q_1 = q_2$
or $q_1+q_2 =1$. For the first case, both $q_1$ and $q_2$ have to be positive
integers since $q_1+q_2 \geq 1$. But, it is not difficult to see that the
states of the form $|m,m\rangle$ with $m$, a positive integer, are all
$Q$-exact. In general, the states of the form $B_0^\epsilon\, \bar{p}_0^n
|m,m\rangle$ where $\epsilon$ = 0 or 1 as before and $n=0,1,2,\ldots$ can
be shown to be $Q$-exact, since $Q$ anticommutes with $B_0$
and commutes with $\bar{p}_0$. Other physical modes $x_0$ and $C_0$ annihilate
the vacuum.

It, therefore, follows that the non-trivial physical states can be obtained
by applying the physical modes on the picture changed vacuum of the form
$|q_1, 1-q_1\rangle$ for $q_1>0$ or $q_1\leq 0$. When $q_1>0$, we
note that the physical modes which do not annihilate the vacuum are $C_0$
and $\bar{p}_0$, but all the states of the form $C_0^\epsilon \bar{p}_0
^n |q_1, 1-q_1\rangle$, for $q_1>0$ can be shown to be cohomologically
trivial except when $q_1 =1$ and $\epsilon = 0$. In that case, the states
lie on the edge of the Kac-table. For other cases, they give combinations
of null vectors and $Q$-exact terms (for concrete examples see [20]).
For $q_1 \leq 0$, the physical modes
which do not annihilate the vacuum are $B_0$ and $x_0$. So, in this case,
states would be of the form
\begineq
B_0^\epsilon\,x_0^n |-m, 1+m\rangle \qquad\qquad {\rm where}\qquad m\geq 0
\endeq
It is easy to see that for $m=0$, $|0, 1\rangle = C_0 |0\rangle$ and so, they
do not generate any new state as they have already been obtained in (29-33).
For $m=1$, the vacuum itself lies on the edge of the Kac-table and
\eqabegin
& &x_0 |-1, 2\rangle\nn\\
&=&:\left[-\frac{i}{2\lambda}\del^2 \phi_M(0) +\frac{1}
{2\lambda}\del^2\phi_L(0) + i\del^2\psi(0) + \frac{1}{2} \del\phi_M(0)
\del\phi_M(0) - \lambda \del\phi_M(0)\del\psi(0)\right.\nn\\
& &\qquad\left. + \frac{1}{2}
\del\phi_L(0)\del\phi_L(0) + i\lambda \del\phi_L(0)
\del\psi(0)\right]\,e^{-\frac{i}{2\lambda}\phi_M(0)+\frac{1}{2\lambda}
\phi_L(0)-i\psi(0)}: |0\rangle\nn \\
&=& -:a(0)\,y(0): |0\rangle
\eqaend
\newpage
Also,
\eqabegin
& & B_0 |-1, 2\rangle\nn\\
&=& :\Big[i\lambda \del\phi_M(0) + \lambda \del\phi_L(0) + i\del\psi(0)
\Big]\,e^{-\frac{i}{2\lambda}\phi_M(0)+\frac{1}{2\lambda}\phi_L(0)}:
|0\rangle\nn\\
&=& y(0)|0\rangle
\eqaend
Here $y(z) = :\big[b(z)c(z) + \lambda \big(i\del\phi_M(z) + \del\phi_L(z)
\big)\big]\,e^{-\frac{i}{2\lambda}\phi_M(z)+\frac{1}{2\lambda}\phi_L(z)}:
$ is the other ground ring generator of $c_M<1$ string theory. In general,
we find,
\eqabegin
x_0^n |-1, 2\rangle &=& -:a(0) y(0) x^{n-1}(0):|0\rangle\\
B_0 x_0^n |-1, 2\rangle &=& :y(0) x^n(0): |0\rangle
\eqaend
For $m=2$, the vacuum itself again lies on the edge of the Kac-table but,
\eqabegin
x_0^n |-2, 3\rangle &=& -\frac{1}{2!}:a(0) y^2(0) x^{n-1}(0): |0\rangle\\
B_0 x_0^n |-2, 3\rangle &=& \frac{1}{2!} y^2(0) x^{n-1}(0): |0\rangle
\eqaend
For a general picture changed vacuum of this type ($m>0$), we have,
\eqabegin
x_0^n |-m, m+1\rangle &=& -\frac{1}{m!} :a(0) y^m(0) x^{n-1}(0): |0\rangle\\
B_0 x_0^n |-m, m+1\rangle &=& \frac{1}{m!} :y^m(0) x^{n-1}(0): |0\rangle
\eqaend
Until now we have seen how to recover all the powers of $x$ and $y$
as well as
the operators multiplied by $a$ in the topological model.
Now we show how the tachyons of $c_M<1$ string theory can be obtained as a
picture changed vacuum. Let us recall from the results of $c_M<1$ string
theory [25] that the tachyons whose matter momenta lie inside the Kac-table,
and whose Liouville momenta satisfy $p_L < Q_L$, can be written in general
as:
\begineq
:w^{-1}(z) x^{p-m'-1}(z) y^{m-1}(z): = :c(z)\,e^{i\alpha_{m,m'}\phi_M(z)
+ i\beta_{m,-m'}\phi_L(z)}:
\endeq
Here $w^{-1}(z) = :c(z)\,e^{i\alpha_{1,p-1}\phi_M(z)+i\beta_{1,-p+1}\phi_L(z)
}:$, $m$ and $m'$ are integers with the restrictions $1\leq m'\leq p-1$;
$1\leq m\leq q-1$ for $(p,q)$ minimal models coupled to 2d gravity. Also,
\eqabegin
\alpha_{m,m'} &=& \frac{1}{2\lambda}(1-m) - \lambda (1-m')\\
\beta_{m,m'} &=& \frac{i}{2\lambda}(1-m) + i\lambda (1-m')
\eqaend
It is easy to check that the picture changed vacuum with $q_1 = (1-m)+2m'
\lambda^2$ and $q_2 = m-2m'\lambda^2$ will have the correct matter and
Liouville momenta of the tachyonic state (51).
We note that the picture charge for the vacuum
associated with the tachyons are fractional and so, none of the modes of
$x$, $\bar{p}$, $B$ and $C$ are well-defined on the picture changed vacua.
Since for these cases $q_1+q_2=1$, the picture
changed vacua themselves are $Q$-invariant but, no new states can be
obtained by applying the modes of $x$, $\bar{p}$, $B$ and $C$ on the vacua.

This, therefore, completes our analysis how to recover the physical states
of ghost number\footnote[1] {Recall that the ghost number of a physical state
equals to the ghost number of the corresponding operator minus one [18].} zero
(tachyons) and $-1$ (ground ring generators) of $c_M<1$
string theory in the associated topological sigma model. Finally we make a
comment about the ghost number $-2$ state $w(0)|0\rangle$, present
in $c_M<1$ string theory [24,25] and whose powers generate the higher ghost
number states. The general form of this state is,
\eqabegin
w(0)|0\rangle &=& :{\cal P} (\del\phi_M, \del\phi_L, b, c)\,e^{i\alpha_{q-1,1}
\phi_M(0)+i\beta_{1,p+1}\phi_L(0)}:|0\rangle\\
{\rm or} &=& :{\cal P} (\del\phi_M, \del\phi_L, b, c)\,e^{i \alpha_{1,p-1}
\phi_M(0)+i\beta_{q+1,1}\phi_L(0)}:|0\rangle
\eqaend
where ${\cal P}$ is a differential polynomial of conformal weight $(p+q-1)$.
The form of $w(0)|0\rangle$ would be quite complicated in general, but for
small $(p,q)$
values it can be calculated with reasonable effort. In fact, for (2,3) model
coupled to gravity, it has the form:
\begineq
w(0)|0\rangle = :\left[2i \del^2\psi(0)+\del\psi(0)\del\psi(0)-\frac{3}{2}
i\del\psi(0)\del\phi_L(0)+{\sqrt 3} \del^2\phi_L(0)\right]\,e^{{\sqrt 3}
\phi_L(0)+i\psi(0)}:|0\rangle
\endeq
where $:e^{i\psi(z)}:=b(z)$. We find that for $(2,3)$ model, this state can
be obtained as,
\eqabegin
w(0)|0\rangle &=& \frac{1}{16}\bigg[23 B(0)\del C(0) - 3 \del B(0) C(0)
+6 x(0) \del \bar{p}(0) + 30 \del x(0) \bar{p}(0)\bigg. \nn\\
& & \left.\qquad -9 \Big(x(0)\bar{p}(0)\Big)^2 + 6 B(0) C(0) x(0) \bar{p}(0)
\right]:|-\frac{3}{2}, \frac{1}{2}\rangle
\eqaend
where $|-\frac{3}{2}, \frac{1}{2}\rangle$ denotes the picture changed vacuum
with picture charge $q_1 = -\frac{3}{2}$ and $q_2 = \frac{1}{2}$. We remark
that since $q_1+q_2 <1$ for this vacuum, it is not $Q$-invariant by itself,
but the whole combination on the right hand side of (57) is $Q$-invariant.
Also,
it is clear that, since the picture charge for this vacuum is fractional, the
individual modes of the fields $B(z), C(z), x(z)$ and $\bar{p}(z)$ are not
well-defined on this vacuum but the composite fields appeared in (57) are
perfectly well-defined. In this sense, $w(0)|0\rangle$ does have a good
description in the topological sigma model we have obtained.

It should be pointed out here that, there are, in fact, two sets of $w$ and
$w^{-1}$ in $c_M<1$ string theory and they are necessary in order to have
a well-defined product among themselves. These two sets can easily be seen
to be related by the symmetry $\lambda \leftrightarrow \frac {1}{2\lambda}$,
$\phi_M \leftrightarrow -\phi_M$; $\lambda \leftrightarrow \frac{1}
{2i\lambda}$, $\phi_M \leftrightarrow \phi_L$ mentioned before. Since the
bosonization (7-10) does not respect this symmetry, we did not get both
sets of
the operators in one particular bosonization. But, we notice that the matter
sector of the second set of $w$ and $w^{-1}$ belongs to the dual of the first
set, so, one would expect them to appear in the dual representation of the
picture changed vacuum. We have checked that this is indeed the case.

To conclude, we have shown that any $(p,q)$ minimal model coupled to gravity
can be regarded as the bosonized form of a constrained topological sigma
model in analogy with the corresponding result for $c_M=1$ string theory.
We have shown that not only the energy--momentum tensor and BRST charge
of these two models are identical under this bosonization but also the
physical states in both these theories are in agreement. Our approach
clarified, as a byproduct, the origin of the twisted $N=2$ superconformal
algebra in $c_M<1$ string theory. There are, however, some subtleties which
we have pointed out. The significance of the ghost number $-2$ state is not
quite clear as it is obtained on a picture changed vacuum which is not
$Q$-invariant. Finally, we mention that, as in the topological gravity
formulation of $c_M<1$ string theory [14], we do not get the restrictions
on the ground ring generators of the
form $x^{p-1}=y^{q-1}=0$. These were imposed in $c_M<1$ string theory
such that the matter momenta of the physical states would lie inside the
Kac-table. There are, however, differences of opinion about this issue.
A different physical state structure has been proposed in ref.[27] for
$c_M<1$ string theory using the descent equations of the double cohomology,
i.e., the usual string BRST cohomology and Felder's BRST cohomology.
In that case,
one does allow all the powers of the ground ring generators
modulo the equivalence relation $x^p \simeq
y^q$, as the physical states. How to implement similar procedure
in the topological sigma model
representation of $c_M<1$ string theory we obtained is not clear to us.
\parbigskip
\parbigskipn
{\bf ACKNOWLEDGEMENTS:}
\parmedskip
We would like to thank A. Fujitsu for providing us his mathematica package
OPEconf.math and ref.[28] which were extensively used for our calculations.
The work of P. M. Ll. is supported by the ``Human Capital and Mobility
Program'' of the European Community and that of S. R. was performed as
part of the research program of the ``Stichting voor Fundamenteel Onderzoek
der Materie''(FOM).
\newpage
\noindent {\bf REFERENCES:}
\parbigskip
\begin{enumerate}
\item S. Mukhi and C. Vafa, \np 407 (1993) 667.
\item P. Horava, \np 386 (1992) 383.
\item O. Aharony, O. Ganor, J. Sonnenschein and S. Yankielowicz,
\pl 305 (1993) 35.
\item D. Ghoshal and S. Mukhi, preprint MRI-PHY/13/93, TIFR/TH/93-62,
hep-th/9312189.
\item A. Hanany, Y. Oz and M. R. Plesser, preprint IASSNS-HEP-94/1,
TAUP-2130-93, WIS-93/123/Dec.-PH, hep-th/9401030.
\item T. Eguchi and H. Kanno, preprint UT-674, hep-th/9404056.
\item H. Ishikawa and M. Kato, preprint UT-Komaba/93-7, hep-th/9304039
(revised Dec. '93) (to appear in Int. Jour. Mod. Phys. A).
\item E. Witten, \np 340 (1990) 281; Surv. Diff. Geom. 1 (1991) 243.
\item K. Li, \np 354 (1991) 711, 725.
\item R. Dijkgraaf, preprint IASSNS-HEP-91/91 and references therein.
\item B. Gato-Rivera and A. Semikhatov, \pl 288 (1992) 295.
\item M. Bershadsky, W. Lerche, D. Nemeschansky and N. Warner, \np
401 (1993) 304.
\item S. Panda and S. Roy, \pl 317 (1993) 533.
\item P. M. Llatas and S. Roy, preprint UG-5/94, hep-th/9406131 (to appear
in Phys. Lett. B).
\item J. Distler, \np 342 (1990) 523.
\item J. M. F. Labastida, M. Pernici and E. Witten, \np 310 (1988)
611.
\item E. Witten, \cmp 118 (1988) 411.
\item E. Witten, \np 373 (1992) 187; E. Witten and B. Zwiebach,
\np 377 (1992) 55.
\item D. Kutasov, E. Martinec and N. Seiberg, \pl 276 (1992) 437.
\item P. M. Llatas and S. Roy, in preparation.
\item R. Dijkgraaf, H. Verlinde and E. Verlinde, preprint PUPT-1217,
IASSNS-HEP-90/80.
\item B. Lian and G. Zuckerman, \pl 254 (1991) 417.
\item P. Bouwknegt, J. McCarthy and K. Pilch, \cmp 145 (1992) 541.
\item H. Kanno and M. H. Sarmadi, \ijmp 9 (1994) 39.
\item S. Panda and S. Roy, \pl 306 (1993) 252.
\item D. Friedan, E. Martinec and S. Shenker, \np 271 (1986) 93.
\item S. Govindarajan, T. Jayaraman and V. John, \np 402 (1993) 118.
\item A. Fujitsu, Computer Physics Communications 79 (1994) 78.
\end{enumerate}
\end{document}